\documentclass[12pt]{article}
\usepackage{latexsym}
\usepackage{amssymb} 

\newcommand{\be}{\begin{equation}}
\newcommand{\ee}{\end{equation}}
\newcommand{\bea}{\begin{eqnarray}}
\newcommand{\eea}{\end{eqnarray}}
\newcommand{\beas}{\begin{eqnarray*}}
\newcommand{\eeas}{\end{eqnarray*}}

\newcommand{\ba}{\begin{array}}
\newcommand{\ea}{\end{array}}

\newcommand{\ie}{{\em i.e.~}}
\newcommand{\eg}{{\em e.g.~}}

\begin{document}


\thispagestyle{empty}

\begin{flushright}
ROM2F/2004/35
\end{flushright}

\vspace{1.5cm}

\begin{center}

{\LARGE {\bf Rationality of the Anomalous Dimensions \\
in ${\cal N}$=4 SYM theory \rule{0pt}{25pt} }} \\
\vspace{1cm} \ {Luigi Genovese
and Yassen  S. Stanev} \\
\vspace{0.6cm} {
{\it Dipartimento di Fisica, \ Universit\`a di Roma \ ``Tor Vergata''}} \\  {{\it I.N.F.N.\ -\ Sezione
di Roma \ ``Tor Vergata''}} \\ {{\it Via della Ricerca  Scientifica, 1}}
\\ {{\it 00133 \ Roma, \ ITALY}}
\end{center}

\vspace{1cm}

\begin{abstract}
We reconsider the general constraints on the perturbative anomalous dimensions 
in conformal invariant QFT and in particular 
in  ${\cal N}=4$ SYM with gauge group $SU(N)$. 
We show that all the perturbative corrections to 
the anomalous dimension of a renormalized gauge invariant
local operator can be written as 
polynomials in its one loop anomalous dimension. 
In the ${\cal N}=4$ SYM theory the coefficients of these polynomials are  
rational functions of the number of colours $N$.
\end{abstract}

\newpage

\setcounter{page}{1}


\section{Introduction and summary of the results}\label{sec:IntroY}

In a conformal invariant quantum field theory (CFT) in which the conformal
operator product expansion (OPE) holds, all $n$-point correlation
functions are completely determined if the following two
ingredients are known:

- the spectrum of the conformal (scale) dimensions $\Delta(g^2)$,
or equivalently the anomalous dimensions $\gamma(g^2) =
\Delta(g^2) - \Delta_0$ of all the operators;

- the OPE structure constants and the normalizations  of
the 3-point functions of the operators\footnote{These two
quantities are related by the normalizations of the 2-point
functions of the operators, which are arbitrary. Only the ratio of
the square of the normalization of the 3-point function and the
product of the normalizations of the 2-point functions has a
invariant meaning.}. The  structure constants determine in
particular also the ``Fusion rules" \ie the  (conformal families of) 
operators that can appear in the product of two given
operators.

Following the  conjectured AdS/CFT correspondence
\cite{ads/cft}, CFT and in particular  ${\cal N}=4$
supersymmetric Yang-Mills (SYM) theory
has been extensively studied in the last several years.
Considerable progress has been achieved both in understanding the
general structure of the theory and in the determination of the
spectrum of the anomalous dimensions. In particular the
complicated structure of $SU(2,2|4)$ supermultiplets has been
understood \cite{dobrevpetkova, ferrarandri} and a classification
of the various shortening conditions was obtained \cite{ferrarashort,
edensokmult, dolanosbornmult}. Non-renormalization theorems for
various types of functions have been proven: for 1/2 BPS operators
\cite{howwestnonrenorm}, as well as extremal \cite{bianchikovacs} and
next-to-extremal correlators \cite{nextoextremal}. The
implications of the superconformal Ward identities have also been
investigated in detail \cite{generalward, dolanosbornpartial, osbornward}.

Explicit calculations of 2-, 3- and 4-point correlation functions
(mostly of protected 1/2 BPS operators) have been performed up to
order $g^4$ \cite{emeri3pt, eden+g2, bianchilog, penati2pt1, eden+g4,  bianchig4, 
penati2pt2, dhoker3pt, gleb+emeri}. 
The simplest 4-point correlation
functions involving  non-protected operators (the lowest component
of the Konishi supermultiplet) have been computed in
\cite{BianchiKonishi, bianchionopmix}. All these computations
confirmed (at least up to order $g^4$) the predicted finiteness of
the correlation functions of gauge invariant operators, and the
resummation of the perturbative logarithms in powers, but
they also demonstrated a complicated mixing pattern for the
operators in the theory.

The spectrum of anomalous dimensions and the related mixing
problem has been investigated  by (essentially) three different
approaches:

- by explicit perturbative calculations of 2-point functions at
order $g^2$, $g^4$ and recently for the Konishi multiplet also at
order $g^6$ \cite{AnselmiKong2, bianchionopmix, plefka++, Constable, gleb+emerisinglet, 
beisert++mix, beisertBMN, burkhardg4, bianchisurprises, burkhardg6};

- by OPE analysis of 4-point functions 
\cite{eden+g2, bianchilog, eden+g4, bianchig4, petkou++, 
dolanosbornpartial, BianchiKonishi,  bianchionopmix, gleb+emeri};

- by explicit diagonalization of the action of the Dilation
generator \cite{minahanzarembo, beisertKstaud, 
beisertcomplete, beisertstaud,  beisertsu23, beiserttesi}. 
This approach, which (in principle) gives all
anomalous dimensions at order $g^2$ \cite{beisertcomplete},
combined with the integrability assumption in the planar limit
($N \rightarrow \infty$) predicts the values of some anomalous
dimensions up to order $g^{12}$.  

To summarize, we have a lot of data at least for the first several
perturbative corrections to the naive scale dimensions of many operators.
Less is known, however, for the general structure of these perturbative 
corrections. 
 
In this paper, rather than performing explicit 
perturbative calculations, we shall reconsider the general constraints 
on the perturbative anomalous dimensions in any CFT and 
in particular in  ${\cal N} = 4$ SYM with gauge group $SU(N)$. 
The main result of our analysis are two properties, which we  call 
 {\it Universality} and {\it Rationality}.
To formulate them we need the following definition: Let $\{ {\cal O}_i \}$, 
$i=1,\dots,d$,  be a complete set of gauge invariant bare operators
which have the same tree level quantum numbers and thus can mix
among themselves. The counterpart of this set, after resolving the mixing
in the interacting CFT, are exactly $d$ renormalized 
operators $\widehat {\cal O}_k$,  $k=1,\dots,d$, with well defined
anomalous dimensions. We define a {\it Class of renormalized operators}
as the set $\{ \widehat {\cal O}_k \}$ which contains all the operators
$\widehat {\cal O}_k$ ,  $k=1,\dots,d$, corresponding to the same set of bare operators 
$\{ {\cal O}_i \}$. It follows that 
each renormalized operator belongs to exactly one 
class\footnote{The partition of the space of operators  
in classes should not be confused 
with the supermultiplet structure in ${\cal N} = 4$ SYM. 
The members of the same class 
of renormalized operators can belong to different 
supermultiplets. Moreover they can be superprimary 
as well as superdescendants.}. We shall prove:

{\it Universality} : In a finite CFT, the
anomalous dimensions of any renormalized
operator $ \widehat {\cal O}_k$ can be written as a polynomial
\be
\gamma_k(g^2) = \sum_{\ell=0}^{d-1} \, w_{\ell} \, (g^2) \left( \gamma_{k}^{(1)} \right)^{\ell} \, .
\label{uni1}
\ee
Here $\gamma_k(g^2)$ is the complete (perturbative) anomalous
dimension of the operator $ \widehat {\cal O}_k$, $\gamma^{(1)}_k$ is its
one loop (order $g^2$) anomalous dimension,
and $d$ is the dimension of the class of renormalized operators 
$\{  \widehat{\cal O}_k \}$
to which $ \widehat {\cal O}_k$ belongs.
The functions 
$ w_{\ell} \, (g^2)$ are universal, \ie they are 
the same for all the operators in a given class of renormalized 
operators. For two different classes of renormalized operators,  these
functions in general  will be different.  

{\it Rationality} : In ${\cal N}=4$ SYM theory with gauge group $SU(N)$
the coefficients $ w_{\ell} \, (g^2)$ in eq.~(\ref{uni1})
will depend also on the number of colours $N$. The functions
$ w_{\ell} \, (g^2,N)$ have a power series 
expansion in $g^2$ with coefficients that are rational functions 
(\ie ratio of polynomials) 
of the number of colours $N$.  In other words
for all $p$  the order $g^{2p}$ 
anomalous dimension of the operator $ \widehat {\cal O}_k$, $\gamma_k^{(p)}$, can be written as  
a polynomial in
$\gamma_{k}^{(1)}$ with universal coefficients, $r^{(p)}_{\ell } (N)$, 
rational in $N$ 
\be
\gamma_k^{(p)} = \sum_{\ell=0}^{d-1} \, r^{(p)}_{\ell } (N) \, \left( \gamma_{k}^{(1)} \right)^{\ell} \, .
\label{rat1}
\ee

Similar considerations apply for the normalizations of the 
$n$-point functions in the theory,
which also can be written as polynomials in the one loop anomalous
dimensions $\gamma^{(1)}$ of the respective operators.

As an explicit illustration of these
considerations we present the rational (in the sense of eq.~(\ref{rat1}))
representation of the order $g^4$ anomalous 
dimensions of the scalar operators of naive scale dimension
$\Delta_{0} = 4$ in the ${\bf 20}^\prime$ representation of the
$SU(4)$ R-symmetry in ${\cal N}=4$ SYM theory. 

\section{ Anomalous dimensions in  conformal field theory }
\label{sec:twopoint}

We shall first recall some relevant general properties of the 
2-point functions in a conformal invariant theory 
 \cite{bianchisurprises}.
Let the scalar\footnote{We choose scalars just for simplicity, the 
case of operators of arbitrary spin  is not essentially different.} 
operators $\widetilde {\cal  O}_i(x,\epsilon)$, with
$i=1,\ldots,d$, be a set of bare  regularized
(by point-splitting with separation $\epsilon$)
operators which can mix among 
themselves, hence they have the same naive dimension $\Delta_0$ \footnote{This,
together with the equality of the spins of the operators, 
is only a necessary condition for mixing, in particular cases 
the operators have to satisfy more conditions, \eg in the case of ${\cal N}=4$ SYM
they have to belong to the same $SU(4)$ representation.}. 
The tilde in $\widetilde {\cal  O}_i(x,\epsilon)$ denotes that these operators are properly
subtracted, as discussed in \cite{bianchisurprises}. 
We want to resolve the mixing problem and find the
corresponding anomalous dimensions. The result of the perturbation
theory calculation will have the form\footnote{Although to simplify the formulae 
we shall write all the relations as depending on only one coupling constant, $g$, the 
generalization to the case of several perturbative coupling constants 
is straightforward.}
\begin{equation}
\langle \widetilde {\cal O}_i(x,\epsilon) \
\widetilde{\cal O}_j^{\dagger}(y,\epsilon)\rangle  =  f_{ij}
\left({\epsilon^2\over (x-y)^2 },g\right) \ {1\over [(x-y)^2]^{\Delta_0}}\, ,
\label{2ptbare}
\end{equation}
where $f_{ij}$ is the non-vanishing (either divergent or finite) 
in the limit $\epsilon \rightarrow 0$ part of the correlator.
It is a hermitian matrix depending on the operator basis we
have chosen. Actually, since complex operators come in pairs with
the same anomalous dimension we can always choose a basis in which
$f_{ij}$ is real and symmetric.

The renormalized operators from the class 
$\{  \widehat{\cal O}_k \}$,  $k=1,\dots,d$,
which have well defined anomalous dimensions $\gamma_k(g^2)$
are linear combinations of the bare operators $\widetilde {\cal O}_j$
\begin{equation}
\widehat {\cal O}_k(x,\mu)  = \sum_{j=1}^{d} Z_{k j}(\epsilon^2 \mu^2,g)
\ {\widetilde {\cal O}}_{j}(x,\epsilon)\, ,
\label{opren}
\end{equation}
where the auxiliary scale $ \mu $ is the subtraction point,
and  $Z$ is the invertible mixing matrix. 
In the basis in which
$f_{ij}$ is real and symmetric $Z$ has the property
$Z^{\dagger} = Z^{T}$, where the superscript $T$ means transposition.
Scale invariance implies that the two-point functions of
$\widehat {\cal O}_k (x,\mu)$ have the form 
\begin{equation}
\langle\widehat{\cal O}_{k} (x,\mu) \ \widehat{\cal
O}_{\ell}^{\dagger}(y,\mu)\rangle  = {\delta_{k \ell} \  M_k(g) \over [(x-y)^2
]^{\Delta_0} [(x-y)^2 \mu^2 ]^{\gamma_k(g^2)}}\, , 
\label{2ptren}
\end{equation}
where we have separated the dependence on the naive and the
anomalous dimension. 
$M_k(g)$ are the {\it finite} normalizations of the 2-point functions
of the renormalized operators~\footnote{Although the 2-point functions can be always 
normalized to one, as we shall see it is convenient to relax this constraint, so we shall 
leave these normalizations as free parameters for the moment.}.

Let us stress that, while $f_{ij}$, $Z_{k j}$ and  $M_k$ can
in general depend on both even and odd powers of the coupling constant $g$,
the physical anomalous dimensions, $\gamma_k$, can only be functions of $g^2$.
Compatibility among  equations (\ref{2ptbare}), (\ref{opren}) and (\ref{2ptren}) 
implies the matrix equation 
\begin{equation}
Z(\epsilon^2 \mu^2,g)\ f\left({\epsilon^2\over (x-y)^2
},g\right)\ Z^{\dagger}(\epsilon^2 \mu^2,g)  = \left[(x-y)^2 \mu^2
\right]^{-\Gamma(g^2)} \, M(g) \, ,\label{renorm1}
\end{equation}
where $\Gamma(g^2)$ and $M(g)$ are the diagonal matrices of anomalous
dimensions and normalizations of the 2-point functions respectively. 
Unitarity implies that $M_k(g)$ are all positive. 
Thus, since there exists a basis in which both $f$ and
$Z$ are real, the anomalous dimensions are also all real.

We write  eq.~(\ref{renorm1}) in two special cases, namely

 - for $\epsilon^2 \mu^2=1$ and $(x-y)^2 \mu^2 = 1 / \rho $ which yields
\begin{equation}
Z(1,g)  \  f(\rho,g) \ Z^{\dagger}(1,g)  =
\rho^{\Gamma(g^2)} \ M(g) \, ,
\label{renorm2}
\end{equation}

- for $\epsilon^2 \mu^2=\rho$ and $(x-y)^2 \mu^2 = 1 $ which yields
\begin{equation}
Z(\rho,g) \   f(\rho,g) \  Z^{\dagger}(\rho,g)  =  M(g) \ .
\label{renorm3}
\end{equation}
It follows that if $Z(1,g)$ is a solution of eq.~(\ref{renorm2}), then
\begin{equation}
  Z(\rho,g)   =  \rho^{-{1 \over 2} \, \Gamma(g^2)} \  Z(1,g)
\label{zl}
\end{equation}
is a  solution of eq.~(\ref{renorm3}). The last relation has a simple intuitive
meaning, one first defines by means of $Z(1,g)$ the operators with well
defined scale dimension, then renormalizes them by the factor
$(\epsilon^2 \mu^2)^{-{1 \over 2}\Gamma(g^2)}$. 
Thus the $\rho$ dependence in $Z(\rho,g)$ factorizes and we have
to solve only eq.~(\ref{renorm2}) for the unknown $Z(1,g)$ and $\Gamma(g^2)$
for a given function $f(\rho,g)$ and a choice of $M(g)$.

Rather than  trying to solve explicitly  these relations (like in \cite{bianchisurprises}),  
in this paper we shall study the implications  of their general structure.
In order to simplify the notation we shall denote $Z(1,g)$ 
by $Z(g)$. Then equation~(\ref{renorm2}) can be written as 
\be
\sum_{k=1}^d Z^{-1}_{ik}(g) \, Z^{-1}_{jk}(g) \, M_{k}(g) \, \rho^{\gamma_k(g^2)} =  f_{ij}(\rho,g) \, .
\label{eq2surpr}
\ee
Introducing the definition 
\be
 A_{ij}^k(g) = Z^{-1}_{i k}(g) \,  Z^{-1}_{j k}(g) \,  M_{k}(g)
\label{eq3surpr}
\ee
and expanding both sides in power series in ${\rm ln}(\rho)$  
\be
 f_{ij}(\rho,g) = \sum_n {\left( {\rm ln}(\rho) \right)^{n} \over n!} \, F_{ij}^{n} (g) \ , 
 \quad \quad
 \rho^{\gamma_k(g^2)} = \sum_n {\left( {\rm ln}(\rho) \right)^{n} \over n!} \,
 \left(\gamma_k(g^2) \right)^{n} \, ,
\label{eq4surpr}
\ee
we get for every $n \geq 0$ 
\be
\sum_{k=1}^d  A_{ij}^k(g) \,\left(\gamma_k(g^2) \right)^{n} =  F_{ij}^{n} (g)\, .
\label{eq5surpr}
\ee
Note that although the range of all three indices $i,j$ and $k$ in the above equation 
is the same (from $1$ to $d$), 
the first two ($i$ and $j$) label the bare regularized operators, while the last one, $k$, 
labels the renormalized operators. 

Let us now specialize to the case of ${\cal N}=4$ SYM with a gauge group $SU(N)$.
All the quantities in the theory will depend also on the number 
of colours $N$. If we take this into account then  equation (\ref{eq5surpr}) becomes
\be
\sum_{k=1}^d  A_{ij}^k(g,N) \, \left(\gamma_k(g^2,N) \right)^{n} =  F_{ij}^{n} (g,N)
\label{srteq1}
\ee
for every $n \geq 0$.

The properties of the renormalized operators are independent of the choice of the 
 basis of  bare regularized operators. 
We shall use this freedom, and choose bare operators 
which do not contain explicit non-rational dependence on $N$. Such a choice indeed always exists, 
for example the pure colour traces which do not contain any explicit $N$ dependence
satisfy this requirement. 
With such a choice of basis of bare operators, 
since the colour contractions can produce only factors rational in $N$, 
the rhs of eq.~(\ref{srteq1}) can be expanded in power series in $g$ as  
\be
  F_{ij}^{n} (g,N) =\sum_{p=0}^{\infty}  g^{2n+p} \ {R_{ij}^{n}}^{(p)} (N) \, , 
\label{srteq1a}
\ee  
where ${R_{ij}^{n}}^{(p)} (N)$ are all rational functions (ratios of polynomials) of $N$.     
Note that even if all ${R_{ij}^{n}}^{(p)}$ are rational, $F_{ij}^{n}$ can still be non-rational
functions of $N$ due to the infinite sum in eq.~(\ref{srteq1a}). 

Expanding in power series in $g$ also the lhs of eq.~(\ref{srteq1}) (and in order to simplify the notation
again suppressing the $N$ dependence) and putting
\bea
 \gamma_k(g^2) &=&  g^2 \gamma_k^{(1)} + g^4 \gamma_k^{(2)} + g^6 \gamma_k^{(3)} + \dots \, , \\
  A_{ij}^k (g) &=&  A_{ij}^{k (0)} + g A_{ij}^{k (1)} + g^2 A_{ij}^{k (2)} + 
 g^3 A_{ij}^{k (3)} + g^4 A_{ij}^{k (4)} + \dots \, ,
\label{srteq1b}
\eea
we get
\bea
&&g^{2n} \sum_k  \left( \gamma_k^{(1)} \right)^n  \,  A_{ij}^{k (0)} + \nonumber \\
&&g^{2n+1} \sum_k  \left( \gamma_k^{(1)} \right)^n \,  A_{ij}^{k (1)} +  \nonumber \\
&&g^{2n+2} \sum_k \left( \gamma_k^{(1)} \right)^{n-1}   \left( 
n \, A_{ij}^{k (0)} \, \gamma_k^{(2)}  +
A_{ij}^{k (2)} \, \gamma_k^{(1)}  \right) + \nonumber \\
&&g^{2n+3} \sum_k \left( \gamma_k^{(1)} \right)^{n-1}   \left( 
n A_{ij}^{k (1)} \gamma_k^{(2)}  +
A_{ij}^{k (3)}  \gamma_k^{(1)}  \right) + \nonumber \\
&&g^{2n+4} \sum_k  \left( \gamma_k^{(1)} \right)^{n-2}   \left( 
{n(n-1) \over 2} \,  A_{ij}^{k (0)} \, \left( \gamma_k^{(2)} \right)^{2}  +
n \, A_{ij}^{k (0)} \, \gamma_k^{(3)}\gamma_k^{(1)}  + \right. \nonumber \\
&& \qquad \qquad \qquad \qquad \qquad 
 \left.  n \, A_{ij}^{k (2)} \, \gamma_k^{(2)} \,  \gamma_k^{(1)}  +
A_{ij}^{k (4)}\,  \left( \gamma_k^{(1)} \right)^2 \right) + \dots \, . 
\label{srteq1c}
\eea
Note that since the terms involving $A_{ij}^{k (p)}$ with $p$ even/odd are always 
multiplied by an even/odd power of $g$ the corresponding equations 
are decoupled. So let us first 
consider the system of equations involving only  $A_{ij}^{k (p)}$ with $p$ even. 
Comparing eq.~(\ref{srteq1a}) and eq.~(\ref{srteq1c}) we find  
\be
\sum_{k=1}^d  \left( \gamma_k^{(1)} \right)^n  \,  A_{ij}^{k (0)} = R_{ij}^{n (0)} \, ,
\label{srteq2}
\ee
\be
\sum_{k=1}^d \left( \gamma_k^{(1)} \right)^{n-1}   \left( 
n \, A_{ij}^{k (0)} \, \gamma_k^{(2)}  +
A_{ij}^{k (2)} \,  \gamma_k^{(1)}  \right) = R_{ij}^{n (2)} \, ,
\label{srteq3}
\ee
\bea
&& \sum_{k=1}^d   \left( \gamma_k^{(1)} \right)^{n-2}   \left( 
{n(n-1) \over 2} \,  A_{ij}^{k (0)} \, \left( \gamma_k^{(2)} \right)^{2}  +
n \, A_{ij}^{k (2)}  \, \gamma_k^{(2)}  \, \gamma_k^{(1)} \  + \right. \nonumber \\
&& \qquad \qquad \qquad \quad 
 \left.    n \, A_{ij}^{k (0)} \, \gamma_k^{(3)} \, \gamma_k^{(1)}+
A_{ij}^{k (4)} \, \left( \gamma_k^{(1)} \right)^2 \right)  = R_{ij}^{n (4)} \, ,
\label{srteq4}
\eea
and so forth.

In order to proceed we shall make an important assumption, namely that the
order $g^2$ corrections to the anomalous dimensions, $\gamma_k^{(1)}$, of the operators
in the same class are  non-degenerate, \ie they are all different. 
We shall return later to the more complicated degenerate case.
 
We shall first analyze the system of equations (\ref{srteq2}) which involves only the 
first corrections to the anomalous dimensions of the operators $\gamma_k^{(1)}$.
Let us define 
\bea
\label{strteq5}
S_0 & = & 1 \, , \nonumber \\
S_1 & = & - \sum_{k_1} \gamma_{k_1}^{(1)} \, , \nonumber \\
S_2 & = &  \sum_{k_1<k_2} \gamma_{k_1}^{(1)} \, \gamma_{k_2}^{(1)} \, , \nonumber \\
S_2 & = &  - \sum_{k_1<k_2<k_3} \gamma_{k_1}^{(1)} \, \gamma_{k_2}^{(1)} \, \gamma_{k_3}^{(1)} \, ,
\nonumber \\
&& \dots \\
S_d & = & (-1)^d \, \gamma_{1}^{(1)} \, \gamma_{2}^{(1)} \dots \gamma_{d}^{(1)} \, .\nonumber
\eea  
Then the system of equations (\ref{srteq2}) implies 
\be
\sum_{\ell=0}^d S_{\ell} \ R_{ij}^{n+d-\ell \ (0)} = 0 \, ,
\label{strteq6}
\ee
for every $n \geq 0$. From this it follows that 

 - all $S_{\ell}$ are rational functions of $N$;
 
 - all $\gamma_{k}^{(1)}$ are roots of a degree $d$ polynomial equation with  
 coefficients rational in $N$ 
\be
P_d(\gamma_{k}^{(1)}) = \sum_{\ell=0}^d S_{\ell} \, \left( \gamma_{k}^{(1)} \right)^{d-\ell} = 0 \, ;
\label{strteq7}
\ee

- an arbitrary (fixed) power of $\gamma_{k}^{(1)}$ can be written as a polynomial
of degree not higher than $d-1$  
in $\gamma_{k}^{(1)}$ with coefficients rational in $N$. Indeed with the help of 
eq.~(\ref{strteq7}) we can express $( \gamma_{k}^{(1)})^d$ and all the higher powers   
in such a form.

We shall need also the following  properties: 

{\bf Property 1}. Any polynomial 
$Q_s(\gamma_{1}^{(1)}, \gamma_{2}^{(1)}, \dots , \gamma_{d}^{(1)})$ 
with coefficients rational in $N$, totaly 
symmetric in its $d$ arguments, 
can be written as a rational function only of $N$. 

Indeed, since $Q_s$ is totally symmetric, it is invariant under the 
permutation of its arguments. Hence it can always be 
expressed as a polynomial (with coefficients rational in $N$) 
in the basis of permutation invariants 
$S_1, \dots, S_d$, defined in eq.~(\ref{strteq5}) (which are also rational functions of $N$).

{\bf Property 2}. If $Q$ is an arbitrary  
rational function (ratio of polynomials) of $\gamma_{k}^{(1)}$, $k=1, \dots ,d$,
with coefficients rational in $N$, 
then it can be written in an equivalent form as 
a polynomial in  $\gamma_{k}^{(1)}$, $k=1, \dots ,d$, with coefficients rational in $N$. 

This follows by noting that by an appropriate multiplication with a polynomial in 
$\gamma_{k}^{(1)}$ we can complete the polynomial in the denominator of $Q$ to a permutation 
invariant, which  due to Property 1. is a rational function only of  $N$. 

{\bf Property 3}. If $Q_1(\gamma_{1}^{(1)};  \gamma_{2}^{(1)}, \dots ,\gamma_{d}^{(1)})$ is 
a  rational function of all its arguments, symmetric in 
$\gamma_{2}^{(1)}, \dots , \gamma_{d}^{(1)}$, with coefficients rational in $N$, 
then it can be written as a polynomial in $\gamma_{1}^{(1)}$ 
(of degree not higher than $d-1$) with coefficients rational in $N$. 

First, the denominator can be eliminated as in the previous case 
without changing the symmetry of $Q_1$. Second, due to the symmetry in 
$\gamma_{2}^{(1)}, \dots , \gamma_{d}^{(1)}$, we can write the 
resulting polynomial in terms of the permutation invariants of 
$\gamma_{2}^{(1)}, \dots , \gamma_{d}^{(1)}$, which in turn can be 
expressed in terms of $S_{\ell}$ of eq.~(\ref{strteq5}) 
and  $\gamma_{1}^{(1)}$. In particular one has 
 \bea
 - \sum_{2 \leq k} \gamma_{k}^{(1)} & = & S_1 + \gamma_{1}^{(1)}  \, , \\
\sum_{2 \leq k_1<k_2} \gamma_{k_1}^{(1)} \, \gamma_{k_2}^{(1)} & = & S_2 - 
\gamma_{1}^{(1)} \, \sum_{2 \leq k} \gamma_{k}^{(1)} 
= S_2 + \gamma_{1}^{(1)} \left( S_1 + \gamma_{1}^{(1)} \right) \, ,
\label{perinv2}
\eea  
and so forth.

After these rather long preliminaries let us return to the system of 
eq.~(\ref{srteq2}). We can solve it for 
$A_{ij}^{k (0)}$ as functions of $\gamma_{1}^{(1)}, \gamma_{2}^{(1)}, \dots , \gamma_{d}^{(1)}$. 
The solution   
\be
A_{ij}^{k (0)} = A_{ij}^{k (0)} \, \left( \gamma_{k}^{(1)}; \, \{ \gamma^{(1)}\}  \diagdown \gamma_{k}^{(1)}
\right) \, 
\ee
is a rational function, symmetric in the $d-1$ variables $\{ \gamma^{(1)}\}  \diagdown \gamma_{k}^{(1)}$.
Thus by Property 3. we can write it in an equivalent form as 
\be
A_{ij}^{k (0)} = \sum_{\ell=0}^{d-1} \, a_{ij}^{(0) \ell } (N) \, \left( \gamma_{k}^{(1)} \right)^{\ell} \, ,
\label{strteq8}
\ee
where $a_{ij}^{(0) \ell }$ are rational  functions of $N$. Note that they do not depend on $k$, \ie on the 
particular operator in the class of renormalized  operators. 

Let us now proceed to the more complicated system of eq.~(\ref{srteq3}), which contains both the first and the second 
corrections to the anomalous dimensions of the operators $\gamma_k^{(1)}$ and $\gamma_k^{(2)}$. We can rewrite it 
in the form 
\be
\sum_k \left( \gamma_k^{(1)} \right)^{n}  \, A_{ij}^{k (2)}  = R_{ij}^{n (2)} - 
n \, \sum_k   
A_{ij}^{k (0)} \, \gamma_k^{(2)} \, \left( \gamma_k^{(1)} \right)^{n-1} = W_{ij}^{n} \ .
\label{srteq3a}
\ee  
Repeating the derivation of eq.~(\ref{strteq6}) from eq.~(\ref{srteq2}) we can exclude all  
$A_{ij}^{k (2)}$ and obtain a system of equations, for every $n \geq 1$, involving only  
 $S_{\ell}$ and $W_{ij}^{n}$
\be
\sum_{\ell=0}^d S_{\ell} \ W_{ij}^{n+d-\ell } = 0 \ .
\label{strteq3b}
\ee
This is a linear system for $\gamma_k^{(2)}$, whose solution (if we use also eq.~(\ref{strteq8}), 
and the assumed non-degeneracy of all $\gamma_k^{(1)}$) has the form
\be
\gamma_k^{(2)} = \gamma_k^{(2)} \, \left(
\gamma_{k}^{(1)}; \, \{ \gamma^{(1)}\}  \diagdown \gamma_{k}^{(1)}
\right) \, ,
\label{strteq9m}
\ee  
which implies for every $k=1, \dots ,d$
\be
\gamma_k^{(2)} = \sum_{\ell=0}^{d-1} \, r^{(2)}_{\ell } (N) \, \left( \gamma_{k}^{(1)} \right)^{\ell} \, .
\label{strteq9}
\ee
Here $r^{(2)}_{\ell } (N)$  are rational functions of $N$, which do not depend on $k$. 
Let us stress that although for demonstrating  this  we made use of a particular basis of bare operators 
(\eg pure colour traces), the rationality property is independent of any choice of basis,
since the anomalous dimensions are   invariant properties of the 
renormalized operators.

Neglecting the $N$-dependence, which is particular for the  ${\cal N}=4 $ SYM theory,  
equation~(\ref{strteq9}) implies that in any CFT the order $g^4$ correction  
to the anomalous dimension of every operator can always be written as a polynomial in the order $g^2$ 
correction  to the anomalous dimension of the same operator. 
The coefficients of the polynomial are the same for different operators belonging to the same class
of renormalized operators.

Inserting eq.~(\ref{strteq9}) into eq.~(\ref{srteq3a}) and solving for $A_{ij}^{k (2)}$ we find also  
\be
A_{ij}^{k (2)} = \sum_{\ell=0}^{d-1} a_{ij}^{(2) \ell } (N) \,  \left( \gamma_{k}^{(1)} \right)^{\ell} \, ,
\label{strteq10}
\ee
where $a_{ij}^{(2) \ell }$ are again rational  functions of $N$, which do not depend on $k$.

We are now ready to solve the system of eq.~(\ref{srteq4}). Substituting in it 
eqs.~(\ref{strteq8}), (\ref{strteq9}) and  (\ref{strteq10}) and bringing the terms 
in the first line to the rhs,
we get 
\be
\sum_k \left( \gamma_k^{(1)} \right)^{n-1}   \left( 
n \, A_{ij}^{k (0)} \, \gamma_k^{(3)}  +
A_{ij}^{k (4)}  \, \gamma_k^{(1)}  \right) = {\widetilde R}_{ij}^{n (4)} \, ,
\label{srteq4a}
\ee
where ${\widetilde R}_{ij}^{n (4)}$ are again rational functions of $N$, because they differ from 
$R_{ij}^{n (4)}$  by a totaly symmetric in $\gamma_k^{(1)}$ ($k=1,\dots ,d$) expression. 
This system  is similar to the system in   
eq.~(\ref{srteq3}), so it is immediate to write its solution, namely 
\be
\gamma_k^{(3)} = \sum_{\ell=0}^{d-1} \, r^{(3)}_{\ell } (N) \, \left( \gamma_{k}^{(1)} \right)^{\ell} \, ,
\label{strteq11}
\ee
and 
\be
A_{ij}^{k (4)} = \sum_{\ell=0}^{d-1} \,  a_{ij}^{(4) \ell } (N) \, \left( \gamma_{k}^{(1)} \right)^{\ell} \, , 
\label{strteq11a}
\ee
with  $r^{(3)}_{\ell }$ and $a_{ij}^{(4) \ell }$ rational functions of $N$.

Proceeding in the same way, and taking into account also the equations 
 for $A_{ij}^{k (p)}$ with $p$ odd implied by 
eqs.~(\ref{srteq1a}) and (\ref{srteq1c}),  we derive 
\be
\gamma_k^{(p)} = \sum_{\ell=0}^{d-1} \, r^{(p)}_{\ell } (N) \, \left( \gamma_{k}^{(1)} \right)^{\ell}
\label{strteq12}
\ee
and 
\be
A_{ij}^{k (p)} = \sum_{\ell=0}^{d-1} \, a_{ij}^{(p) \ell } (N) \, \left( \gamma_{k}^{(1)} \right)^{\ell}
\label{strteq13}
\ee
for every $i,j,k$ and $p$ with coefficients $r^{(p)}_{\ell }$ and $a_{ij}^{(p) \ell }$ rational in $N$ and independent of $k$.

It  follows, from eq.~(\ref{strteq12}), that in any CFT any fixed 
order correction to (and hence also the total perturbative) anomalous dimension of every operator can always be 
written as a polynomial in the order $g^2$ 
correction  to the anomalous dimension of the same operator 
\be
\gamma_k(g^2) = \sum_{\ell=0}^{d-1} \, w_{\ell} \, (g^2) \left( \gamma_{k}^{(1)} \right)^{\ell} \, .
\label{strteq14}
\ee
The coefficients of the polynomial, $ w_{\ell} \ (g^2)$, are universal, 
 since they do not depend on $k$ and hence are the same for different 
operators belonging to the same class of renormalized operators.
In particular, in ${\cal N}=4$ SYM, the functions $ w_{\ell} \ (g^2)$ have a power series 
expansion in $g^2$ with coefficients $r^{(p)}_{\ell }$ rational in $N$. 

So far we considered the generic case when the order $g^2$ anomalous dimensions are 
just roots of a degree $d$ polynomial with coefficients rational in $N$  (see eq.~(\ref{strteq7})).
It often happens that this polynomial can be factorized  as a product of two (or more) polynomials,
 with  coefficients still rational in $N$,   
\be
P_d(\gamma_{k}^{(1)}) = P_{d_1}(\gamma_{k}^{(1)}) \  P_{d_2}(\gamma_{k}^{(1)}) =  0 \, .
\label{strteq7n}
\ee
In this case the class of operators naturally splits into two subclasses, one containing $d_1$,
the other $d_2 = d-d_1$ operators. All the statements about the permutation invariants and 
their $N$ dependence are valid separately for each of the two subclasses, so we can write 
eq.~(\ref{strteq14}) in the following equivalent form
\bea
\gamma_{k_1}(g^2) &=& \sum_{\ell=0}^{d_1-1} u_{\ell} \ (g^2) \left( \gamma_{k_1}^{(1)} \right)^{\ell} \, , \nonumber \\
\gamma_{k_2}(g^2) &=& \sum_{\ell=0}^{d_2-1} v_{\ell} \ (g^2) \left( \gamma_{k_2}^{(1)} \right)^{\ell} \, ,
\label{strteq14a}
\eea
where $\gamma_{k_j}^{(1)}$, $j=1,2$ are the roots of  $P_{d_j}(\gamma_{k_j}^{(1)})=0$ respectively. 
We have decreased the degree of the polynomials in this expressions, but at the price of using 
different functions $u_{\ell}$ and $v_{\ell}$ for the two subclasses of operators. Note that the number of 
variables is the same for both representations since $d=d_1+d_2$. Moreover, if the 
functions $ w_{\ell} \ (g^2)$ admit a power series 
expansion in $g^2$ with coefficients rational in $N$, so will the functions $u_{\ell}$ and $v_{\ell}$.  
As a consequence of eqs.~(\ref{strteq7n}) and~(\ref{strteq14a}),  
$\gamma_{k_j}^{(p)}$ at all orders 
in perturbation theory will be given by the roots of a polynomial 
of degree not higher than $d_j$. Hence the factorization of the equation for 
$\gamma^{(1)}$ implies the factorization of the equation for all $\gamma^{(p)}$. Thus the 
dimensions of the closed (in the sense of eq.~(\ref{strteq7n})) one loop subclasses are 
preserved to higher loops. The generalization to more than two factors in eq.~(\ref{strteq7n}) 
is straightforward.

The factorization of the polynomials defining $\gamma^{(1)}$ (like in eq.~(\ref{strteq7n}))
is of particular importance in  ${\cal N}=4 $ SYM theory.
The reason is that, on the one hand the operators in  ${\cal N}=4 $ SYM are organized 
in large supermultiplets with typical number of components of the order of $2^{16}$, all with the  
same anomalous dimension $\gamma(g^2)$. On the other hand, the relevant quantum numbers for the 
resolution of the mixing problem considered in this paper are the naive scale dimension 
$\Delta_0$ and the spin $s$ of the operators and the $SU(4)$ representation 
to which they belong. Hence, in general, the set  of $d$ operators we start with 
may contain both operators which belong to {\it similar}  supermultiplets 
(with the same quantum numbers of the naive, for $g=0$, lowest component 
and the same number of components) 
and  to {\it essentially different}  supermultiplets 
(with different quantum numbers of the 
naive lowest components and/or different number of components).
It turns out that the factorization properties of the polynomial $P_d$  
defining $\gamma^{(1)}$ describe exactly this structure. 
To be more precise, if the order $g^2$ corrections to the anomalous dimensions 
of two operators are roots of a non-factorizable (with coefficients rational in $N$) polynomial, 
then these two operators belong to similar supermultiplets.
The proof follows by noting that if two supermultiplets are essentially different 
then there exists at least one component which belongs to only one, say the first,  of them.
Let us consider the class of renormalized operators corresponding to 
such a component. It follows that this class will contain an operator
 belonging to the first supermultiplet, but will not contain
an operator  belonging to the second supermultiplet.  
This contradicts our assumption that the polynomial is non-factorizable,   
and hence has as roots always both anomalous dimensions.
In other words, if two operators belong to essentially different supermultiplets
then their order $g^2$ anomalous dimensions will be roots of different factors 
of the polynomial in eq.~(\ref{strteq7n}). Hence we can identify the subclasses of 
renormalized operators with the families of essentially different supermultiplets. The functions, 
$u_{\ell} \ (g^2)$ and $v_{\ell} \ (g^2)$, in eqs.~(\ref{strteq14a}) will be universal 
for all the similar supermultiplets within each family.
Whether the inverse is also true, \ie if the factorization of the polynomial 
implies that the respective supermultiplets are essentially different 
is an open challenging problem.  

An important particular case of factorization is when the order $g^2$ anomalous dimension of 
some operator $\widehat {\cal O}_k$, 
$\gamma_{k}^{(1)}$, is a rational function of $N$ (this corresponds to some $d_j=1$), 
as \eg for the components of the Konishi supermultiplet ${\cal K}$ in the ${\cal N}=4 $ SYM theory. 
Then it follows that for such an operator all perturbative corrections, $\gamma_{k}^{(p)}$, will  
be  rational functions of $N$.
Still the complete anomalous dimension, being an infinite series, may not share this property.

Before proceeding, let us briefly comment also on the degeneracy problem.
So far we have assumed that the
order $g^2$ corrections to the anomalous dimensions of the operators, $\gamma_k^{(1)}$,
are non-degenerate, \ie they are all different. This assumption is essential in deriving 
the unique representations for the higher order anomalous dimensions (see \eg eq.~(\ref{strteq12})). 
Indeed, if two order $g^2$  anomalous dimensions, say  $\gamma_{k_1}^{(1)}$ and 
$\gamma_{k_2}^{(1)}$,  are equal 
then the determinant of the coefficients in eqs.~(\ref{srteq2})~-~(\ref{srteq4}) is zero.  
There are two distinct types of degeneracy. On the one hand there are 
operators which have exactly the same anomalous dimension to all orders in perturbation theory
\eg since they belong to the same supermultiplet.
There is no way to lift this degeneracy by considering only the 2-point functions and  
one has to take into account also some 3-point functions to distinguish such operators. 
On the other hand it may happen\footnote{Although we are not aware of any explicit 
example of this kind in ${\cal N}=4 $ SYM.} 
that the degeneracy is removed at some higher order in perturbation theory. 
That is, there exists some $q$ such that  
$\gamma_{k_1}^{(q)} \neq \gamma_{k_2}^{(q)}$ (more precisely we want that all
 the $q$-th  order anomalous dimensions are not degenerate). 
In this case, repeating all the derivations, one can  show that all the above formulae 
remain valid if we  replace  $\gamma_{k}^{(1)}$ by $\gamma_{k}^{(q)}$, hence all anomalous dimensions 
can be written as polynomials in the non-degenerate 
order $g^{2q}$ anomalous dimensions  $\gamma_{k}^{(q)}$ 
(with coefficients independent of $k$ and rational in $N$). 
We are convinced that it is more efficient to solve the degeneracy problem case by case, 
depending on the properties of the particular operators at hand, rather than to develop a general
prescription. Thus in the rest of the paper we shall treat again only the non-degenerate case.  

\section{Normalizations of the correlation functions}\label{sec:Normalizations}

Given the  universality and   rationality  properties of the anomalous dimensions 
in  eqs.~(\ref{strteq12}), (\ref{strteq14}),   a natural question arises: 
Are there, and if yes under which assumptions,  
similar formulae for the normalizations of the 2-point functions
of the renormalized operators, $M_k(g)$, which enter eq.~(\ref{2ptren})?
In other words, can we normalize the renormalized operators in such a way that
\be
M_k(g) = \sum_{\ell=0}^{d-1} m_{\ell} \, (g,N) \, \left( \gamma_{k}^{(1)} \right)^{\ell} \, ,
\label{normrat}
\ee
where $m_{\ell} \ (g,N)$ are  functions independent of $k$ which admit a power series 
expansion in $g$ with coefficients rational in $N$. 
A sufficient condition is the existence of a rational mixing matrix
\be
Z^{R}_{ki}(g) = \sum_{\ell=0}^{d-1} z_i^{\ell} \, (g,N) \, \left( \gamma_{k}^{(1)} \right)^{\ell} \, ,
\label{ZR}
\ee
such that $z_i^{\ell} \ (g,N) $ are   independent of $k$ functions which admit a power series 
expansion in $g$ with  coefficients rational in $N$. Indeed eq.~(\ref{ZR}), together with 
equations~(\ref{eq3surpr}) and~(\ref{strteq13}), imply eq.~(\ref{normrat}).
Let $Z^{R}$ be a matrix of, rational in $N$, eigenvectors of 
$\left( F^0 (g) \right)^{-1}  \left( F^1 (g) \right)$, such that  
\be
Z^{R}(g) \, \left[ \left( F^0 (g) \right)^{-1}  \,  F^1 (g) \right] \,  
\left( Z^{R} (g) \right)^{-1} = \Gamma(g^2) \, ,
\label{eqx}
\ee
with  $F^0(g)$ and $F^1(g)$  defined in eq.~(\ref{eq4surpr}). 
Such a matrix always exists, since by 
expanding in power series in $g$ the eigenvector condition
and using the rationality in $N$
of $F^0$, $F^1$ and $\Gamma$ one proves  the existence of  
 eigenvectors rational in $N$.
The polynomial form of eq.~(\ref{ZR}), 
and hence also of eq.~(\ref{normrat}), then follows.
On the other hand, $Z^{R}(g)$ diagonalizes simultaneously $F^0$ and  $F^1$ 
as required for a mixing matrix
\bea
Z^{R}(g) \, F^0(g) \, \left( Z^{R} (g) \right)^{\dagger} &=& M(g) \, , \nonumber \\
Z^{R}(g) \, F^1(g) \,  \left( Z^{R} (g) \right)^{\dagger} &=& M(g)  \, \Gamma(g^2) \, .
\label{eqz}
\eea 
In fact there exists a (unique) unitary mixing matrix $Z$ \cite{bianchisurprises} which 
satisfies eqs.~(\ref{eqz}) 
with $M(g)=1$, and thus  also eq.~(\ref{eqx}). Then it follows that $Z^R$ and $Z$
can be related by a real diagonal matrix $D(g)$, such that $Z^R(g)  = D(g) Z(g)$, 
which in turn implies  eqs.~(\ref{eqz}) with $M(g)=D^2(g)$.
Note that there is the residual freedom of multiplying from the left the matrix
$Z^R(g)$  by a diagonal matrix with entries polynomial in $\gamma_{k}^{(1)}$ and 
rational in $N$. Such a transformation 
preserves the polynomial structure of both eqs.~(\ref{ZR}) and~(\ref{normrat})
but modifies the values of $M_k(g)$.
However in general it is not possible to get $M(g)=1$ by such a transformation,
hence the standard unitary mixing matrix $Z$ of \cite{bianchisurprises} has 
entries non-rational in $N$.
This explains also why we did not impose the standard normalization condition on the 2-point 
functions in eq.~(\ref{2ptren}).
 
The generalization to the case of higher point functions is as follows.
The normalization, $C_{k_1 k_2 k_3}$, 
of the 3-point function 
$\langle \widehat{\cal O}_{k_1} (x_1) \ \widehat{\cal O}_{k_2} (x_2) \ \widehat{\cal O}_{k_3} (x_3)\rangle$  
has the following representation
\be
C_{k_1 k_2 k_3} (g) = \sum_{\ell_1 , \ell_2 , \ell_3} \, c^{\ell_1 \ell_2 \ell_3} \, (g) \, 
\left( \gamma_{k_1}^{(1)} \right)^{\ell_1} \,
\left( \gamma_{k_2}^{(1)} \right)^{\ell_2} \, \left( \gamma_{k_3}^{(1)} \right)^{\ell_3}  \, .
\label{trip1}
\ee
A similar expression holds also for the OPE coefficients defined as
\be
{C_{k_1 k_2}}^{k_3} (g) = {C_{k_1 k_2 k_3} (g) \over  M_{k_3}(g) }  \, .
\label{opec1}
\ee
Both these quantities will depend on the particular normalizations of the 2-point functions
of the respective operators, $M_{k}(g)$, and thus have no invariant meaning.
In fact the only physical quantities, 
that are independent on any normalization choices, are the ratios
\be
T_{k_1 k_2 k_3} (g^2) = { \left( C_{k_1 k_2 k_3} (g) \right)^2  \over  M_{k_1}(g) \, M_{k_2}(g) \, M_{k_3}(g) }  \, .
\label{tripr1}
\ee
Note that, unlike $C_{k_1 k_2 k_3}(g)$  and ${C_{k_1 k_2}}^{k_3} (g)$ which may depend both on even
and odd powers of $g$, $T_{k_1 k_2 k_3} (g^2)$ is a function of $g^2$ only.
Combining eqs.~(\ref{trip1}) and (\ref{normrat}) we find 
\be
T_{k_1 k_2 k_3} (g^2) = \sum_{\ell_1 , \ell_2 , \ell_3} \, t^{\ell_1 \ell_2 \ell_3} \, (g^2 )\, 
\left( \gamma_{k_1}^{(1)} \right)^{\ell_1} \,
\left( \gamma_{k_2}^{(1)} \right)^{\ell_2} \, \left( \gamma_{k_3}^{(1)} \right)^{\ell_3}  \, ,
\label{tripr2}
\ee
where the functions $t^{\ell_1 \ell_2 \ell_3} \, (g^2) $ do not depend on the choice of
the operators in the three classes of renormalized  operators. In particular, in 
${\cal N} =4$ SYM, the functions $t^{\ell_1 \ell_2 \ell_3} \, (g^2, N)$ 
will have a power series expansion in $g^2$ with  coefficients rational in $N$.
Given the 2- and 3-point functions all higher $n$-point functions can be obtained by the OPE.

\section{Conclusions}\label{sec:Conclusions}
In the framework of CFT 
we derived a representation for all the perturbative corrections to the anomalous dimension
of  any given gauge invariant operator as polynomials in its one loop anomalous dimension,
eq.~(\ref{uni1}).
In the case of ${\cal N} =4$ SYM with gauge group $SU(N)$ we have proven that the 
coefficients of these polynomials are rational functions 
of the number of colours $N$, (see eq.~(\ref{rat1})). 

Since our considerations do not modify
the number of unknown functions, it might seem that this is  
a completely equivalent representation
but, as we shall argue, in ${\cal N} =4$ SYM this is not the case.
The reason is twofold. On the one hand, due to the rationality property of eq.~(\ref{rat1}),
at each given order in perturbation theory,
we express $d$ arbitrary functions of $N$, $\gamma_k^{(p)}(N)$, in terms of   
the same number rational functions of $N$, $r^{(p)}_{\ell }(N)$. 
This has also an important technical implication, since in this way we are able to reconstruct 
the exact analytic form of the anomalous dimensions from solutions which are 
necessarily numerical for a large number of operators. 
On the other hand, the universality property  of eq.~(\ref{uni1}), 
combined with the factorization properties in eqs.~(\ref{strteq7n}),~(\ref{strteq14a}) implies that 
 the functions $w_{\ell} \, (g^2)$ are universal for all the operators 
which belong to a family of similar supermultiplets 
(with the same quantum numbers of the naive lowest component and the same number of components). 
In other words the naive scale dimension 
$\Delta_0$, the spin $s$ and the $SU(4)$ representation 
of the lowest component determine the 
whole set of functions $\{ w_{\ell} \, (g^2)\}$.
Hence to compute the perturbative  
anomalous dimension of any operator in the theory, $\gamma_k(g^2)$,  
in principle it is sufficient to specify its one loop anomalous dimension, $\gamma_k^{(1)}$, 
and the family of similar supermultiplets to which the operator belongs.

This suggests that it should be possible to obtain along these lines  
general formulae for the anomalous dimensions in ${\cal N}=4$ SYM theory.
As an illustration that, if it exists, such a representation is far from obvious we  
shall write down the polynomial form of the order $g^4$ anomalous 
dimensions of the scalar operators of naive scale dimension
$\Delta_{0} = 4$ in the ${\bf 20}^\prime$ representation of the
$SU(4)$ R-symmetry. 
There are four superprimary operators of this kind \cite{bianchionopmix}.  One is protected, 
while the order $g^4$ anomalous dimensions of the other three are given 
by\footnote{These anomalous dimensions, in a different but equivalent form, have been first computed
in \cite{burkhardg4}.} 
\bea
\gamma_k^{(2)} =
\frac{ N^2}{(4 \pi^2)^2}\frac{1}{4 (N^6 + 116 N^4 - 1180 N^2 + 800)} \Bigl[ 
(N^6 - 188 N^4 + 1596 N^2 -1040) \eta_k^2  \; -\Bigr. \nonumber \\
 \Bigl. 2 (5 N^6 -133 N^4 +532 N^2 -220) \eta_k + 
(3 N^2 -2) (3 N^4 - 112 N^2 +500) \Bigr]\;,
\label{gammag4}
\eea
where $k=1,2,3$ and $\eta_k ={\gamma_k^{(1)}}/{(\frac{ N}{4 \pi^2})}$  are the roots of the cubic equation 
\begin{equation}
N^2 \eta^3 -8\,N^2 \eta^2+10(2\,N^2-1)\eta-5(3\,N^2-2)=0\;.
\label{gammag2}
\end{equation}
The details of the calculation, as well as
the expressions for the normalizations of the 3-point functions and the OPE coefficients 
involving these operators will be presented elsewhere \cite{noiinpreparation}. 


\section*{Acknowledgements }

It is a pleasure to thank  M.~D'Alessandro, M.~Bianchi  and especially 
G.~C.~Rossi for numerous discussions and for comments on the manuscript. 
This work was supported in part by INFN, by the MIUR-COFIN contract
2003-023852,
by the EU contracts MRTN-CT-2004-503369 and MRTN-CT-2004-512194, by the INTAS
contract 03-51-6346, and by the NATO grant PST.CLG.978785.




\end{document}